\documentclass[prl,letterpaper,twocolumn,showpacs,superscriptaddress,amsmath,amsfonts,amssymb,preprintnumbers]{revtex4}
\pdfoutput=1
\usepackage[T1]{fontenc}\usepackage[latin1]{inputenc}
\usepackage{dcolumn,graphicx,color,booktabs,microtype,afterpage} 
\usepackage[charter]{mathdesign}

\usepackage[colorlinks,plainpages=false,linkcolor=blue,urlcolor=blue,citecolor=black,pdfpagemode=UseNone,pdfstartview=FitBH]{hyperref}

\begin{document}

\title{Direct observation of superconducting vortex clusters pinned by a periodic array of magnetic dots in ferromagnetic/superconducting hybrid structures.}

\author{T.~Shapoval{\large\hyperref[CorrAuthor]{*}}}
\affiliation{IFW Dresden, Institute for Metallic Materials, P.\,O.\,Box 270116, D-01171 Dresden, Germany.}

\author{V.~Metlushko}
\affiliation{Department of Electrical and Computer Engineering, University of Illinois at Chicago, Illinois 60612, USA.}

\author{M.~Wolf}\author{B.~Holzapfel}\author{V.~Neu}\author{L.~Schultz}
\affiliation{IFW Dresden, Institute for Metallic Materials, P.\,O.\,Box 270116, D-01171 Dresden, Germany.}

\begin{abstract}
Strong pinning of superconducting flux quanta by a square array of 1\,$\mu$m-sized ferromagnetic dots in a magnetic-vortex state was visualized by low-temperature magnetic force microscopy (LT-MFM). A direct correlation of the superconducting flux lines with the position of the dots was derived. It was found, that the superconducting vortices which are preferably located on top of the Py dots experience stronger pinning forces as compared to the pinning force in the pure Nb film. This pinning exceeds the repulsive interaction between the superconducting vortices and allows vortex clusters to be located at each dot. Our microscopic studies are consistent with global magnetoresistance measurements on the hybrid structures, but suggest a modified picture of the pinning mechanism. 
\end{abstract}

\pacs{\vspace{-0.2em}74.70.-b 74.78.-w 74.25.Qt 68.37.Rt\vspace{-0.5em}}

\maketitle
Controlling the distribution of magnetic flux quanta (superconducting vortices) in superconducting materials by introducing artificial pinning centers is a challenge, both in basic and in applied research. Due to the presence of the natural point disorder (e.g. grain and intergrain pinning) in superconducting thin films superconducting vortices form a weakly disordered Abrikosov lattice~\cite{Abr57, Vol02}, so called topologically ordered Bragg glass~\cite{Gia97}. In the last decade a variety of studies has been performed to investigate the influence of different artificial pinning centers on the superconducting properties of thin films~\cite{Mos96, Fol07, RefA, Ala09, RefC}. On the one hand, randomly distributed defects act as strong local pinning centers which significantly improve the in-field critical parameters of superconducting films~\cite{Fol07}, on the other hand, ordered pinning potentials give rise to collective pinning mechanisms and thus lead to commensurate pinning effects~\cite{Mos96, RefA, Pat07}. In comparison to simple structurally ordered pinning sites, magnetic pinning centers provide additional degrees of freedom, which lead to several pronounced effects, such as domain-wall superconductivity, field induced superconductivity, proximity effect, magnetostatic interaction, and local suppression of superconductivity by strong out-of-plane field components~\cite{Ala09, RefC}, some of which can be used to tune vortex dynamics by rectifying vortex motion~\cite{Sil07}. 

In this work, the interplay between superconducting and magnetic vortices in ferromagnetic/superconducting (FM/SC) hybrid structures with well-controlled lateral dimension is visualized by low-temperature magnetic force microscopy (LT-MFM) using a commercial scanning probe microscope (\textit{Omicron Cryogenic SFM})~\cite{Sha07}. The MFM cantilever (Nanoworld MFMR) possesses a force constant of about 2.8\,N/m and a resonance frequency of about 80\,kHz. The measured shift of the cantilever's resonance frequency  $\Delta f$ is proportional to the derivative of the $z$ component of the force that acts between the tip and the sample at a given scanning distance above the surface~\cite{Alb91}. 

\begin{figure}[t]
	\centering
		\includegraphics[width=\columnwidth]{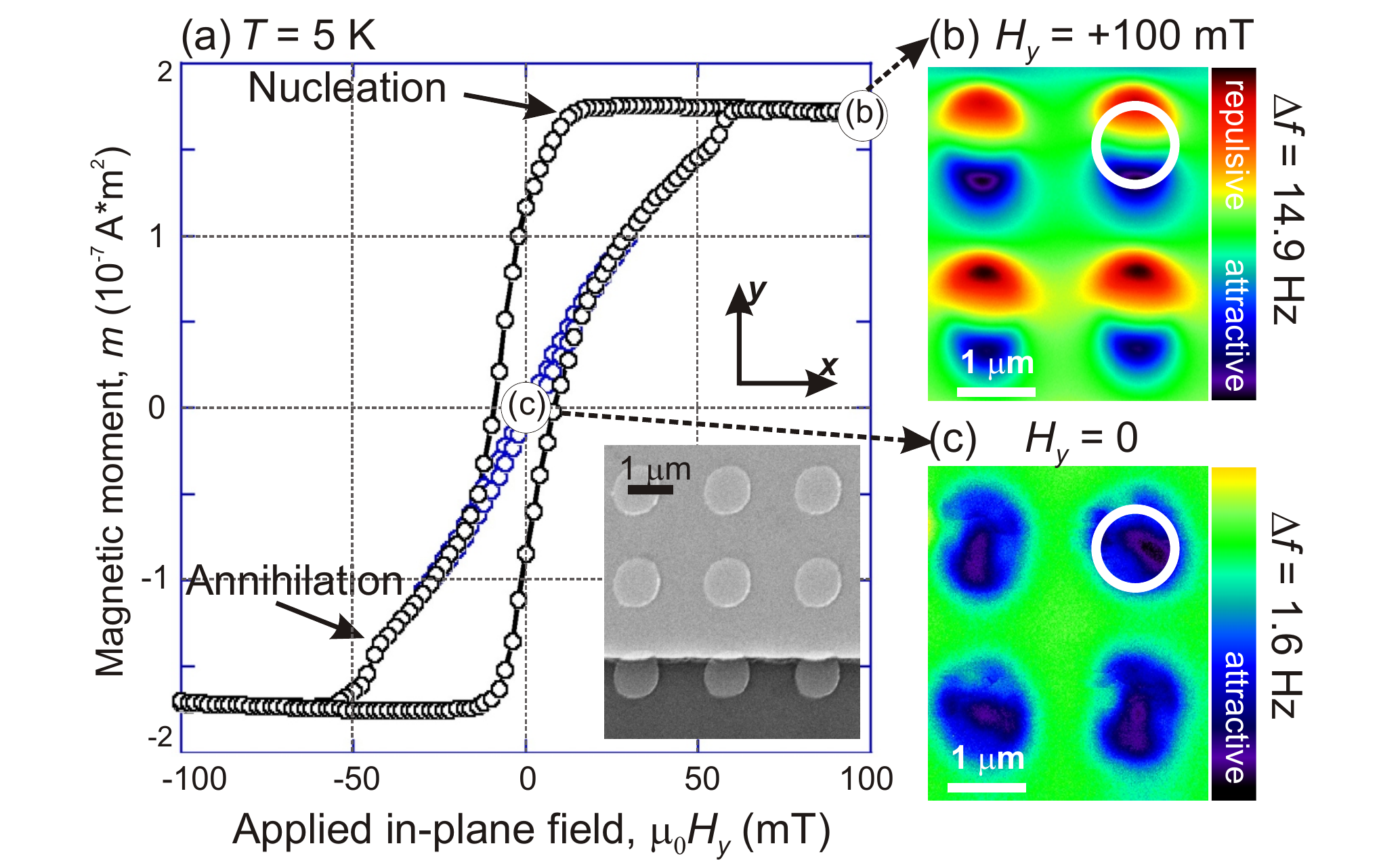}
		\caption{(a) Magnetic hysteresis loop shows magnetic vortex behavior with vortex nucleation and annihilation fields. Within the inner loop, between $-30$\,mT and $+30$\,mT, the magnetization is reversible. The inset shows a SEM image of Py dots covered with a 100\,nm thick Nb layer. The in-plane field $H_y$ was varied along the hysteresis loop starting from saturation at $+100$\,mT (b), through applying a negative field less than the magnetic-vortex annihilation field ($-25$\,mT) to the magnetic-vortex state at zero field (c). Color bars give the measured $\Delta f$ signal which strongly differs between the saturated and the vortex state. A small out-of-plane field of $+10$\,mT was permanently applied to insure a positive polarity of the magnetic vortex. Scanning distance was 75\,nm, $T = 14.6$\,K. The white circles represent the location of the Py dot.}
\label{fig:hysteresis}
\end{figure}

The following hybrid structure is studied: a square array of permalloy (Py = Ni$_{80}$Fe$_{20}$) dots with 1\,$\mu$m diameter, 25\,nm height and 2\,$\mu$m periodicity was prepared on a Si (100) substrate using standard e-beam lithography, e-beam evaporation, and lift-off processes; a 100\, nm thick superconducting niobium (Nb) film ($T_c = 8.32$\,K) was deposited on top of the Py dot array by sputter deposition~\cite{Hof08}. A SEM scan of this Py/Nb hybrid structure is shown in Fig.\,\ref{fig:hysteresis}\,(a)\,(inset).

Depending on their shape and aspect ratio, ferromagnetic dots can be in different magnetization states such as multidomain, single domain or, for the circular dots, a magnetic-vortex state becomes energetically stable at remanence~\cite{Sch00}. Here, the magnetization curls continuously around the center while staying purely in-plane in a large area of the dot and turns perpendicular to the surface in the center of the dot creating a small magnetization swirl~\cite{Wac02, Hub08}. This swirl, also called a magnetic-vortex core, has either positive or negative polarity of its out-of-plane stray field and has a maximum width of $5 l_\text{ex} \approx 25$\,nm for the present geometry, with  $l_\text{{ex}}=\sqrt{A/K_d}$ being the exchange length, where $A$ is the material specific exchange stiffness constant and $K_d$ is the stray field energy constant~\cite{Hub08}. 

Hoffmann \textsl{et al.} have reported a clear correlation between a strong drop in the resistivity curve of the SC Nb film and the  magnetic-vortex state of the underlying Py dots, which was shown to be independent of the polarity of the magnetic-vortex core~\cite{Hof08}. In the present work, local imaging was applied to look deeper into the nature of this enhanced pinning, providing a direct determination of the preferable locations of SC vortices. 

The magnetic in-plane hysteresis loop [Fig.\,\ref{fig:hysteresis}\,(a)] of the Py array measured at 5\,K using a superconducting quantum interference device (SQUID) clearly reveals magnetic vortex behavior with vortex nucleation and annihilation fields. For the inner loop in the field range from $-30$\,mT to $+30$\,mT, the magnetization process occurs only by vortex propagation and, thus, is reversible (vortex branch).

To reach the magnetic-vortex configuration in the Py array, the sample was cooled in the microscope down to 40\,K. An in-plane field of $+100$\,mT was applied along the positive $y$ direction ($H_y$) to fully saturate the dots, and the sample was further cooled down to 14.6\,K. The MFM scan at a tip-sample distance of 75\,nm shows four saturated Py dots [Fig.\,\ref{fig:hysteresis}\,(b)]. Decreasing the field to $-25$\,mT ensures that in most of the dots a magnetic vortex is nucleated. Going back to zero along the vortex branch brings the dots into the symmetric magnetic-vortex state imaged in Fig.\,\ref{fig:hysteresis}\,(c). Magnetic vortices, generated in such a way, will have random polarity (i.e. out-of-plane magnetization components pointing randomly up or down)~\cite{Vil08}. To set a defined polarity, a small positive out-of-plane field ($+10$\,mT) was applied to the sample during the above-described field sequence. Thus, the magnetic-vortex core and the MFM tip, which is magnetized in positive $z$-direction, experience an attractive interaction that shows up as a dark contrast in the center of the dot [Fig.\,\ref{fig:hysteresis}\,(c)]. 

After reaching the magnetic-vortex state of the Py dots the sample was repeatedly cooled down to a temperature below $T_c$ of the Nb film ($T = 6.1$\,K $\approx 72\%$ $T_c$) in perpendicular fields $H_{\text{applied}}= +0.5$\,mT, 0, $-0.5$\,mT and $-1$\,mT that are close to the matching fields for this hybrid structure. The matching field $H_m$ is a field that ensures an integer number $m$ of vortices per unit area $S$ of the dot array: $H_m = m \Phi_{0}/S$, with $\Phi_0=h/2e$ being the magnetic flux quantum and $S=4$\,$\mu$m$^2$~\cite{Mos96}.

\begin{figure}[t]
	\centering
		\includegraphics[width=\columnwidth]{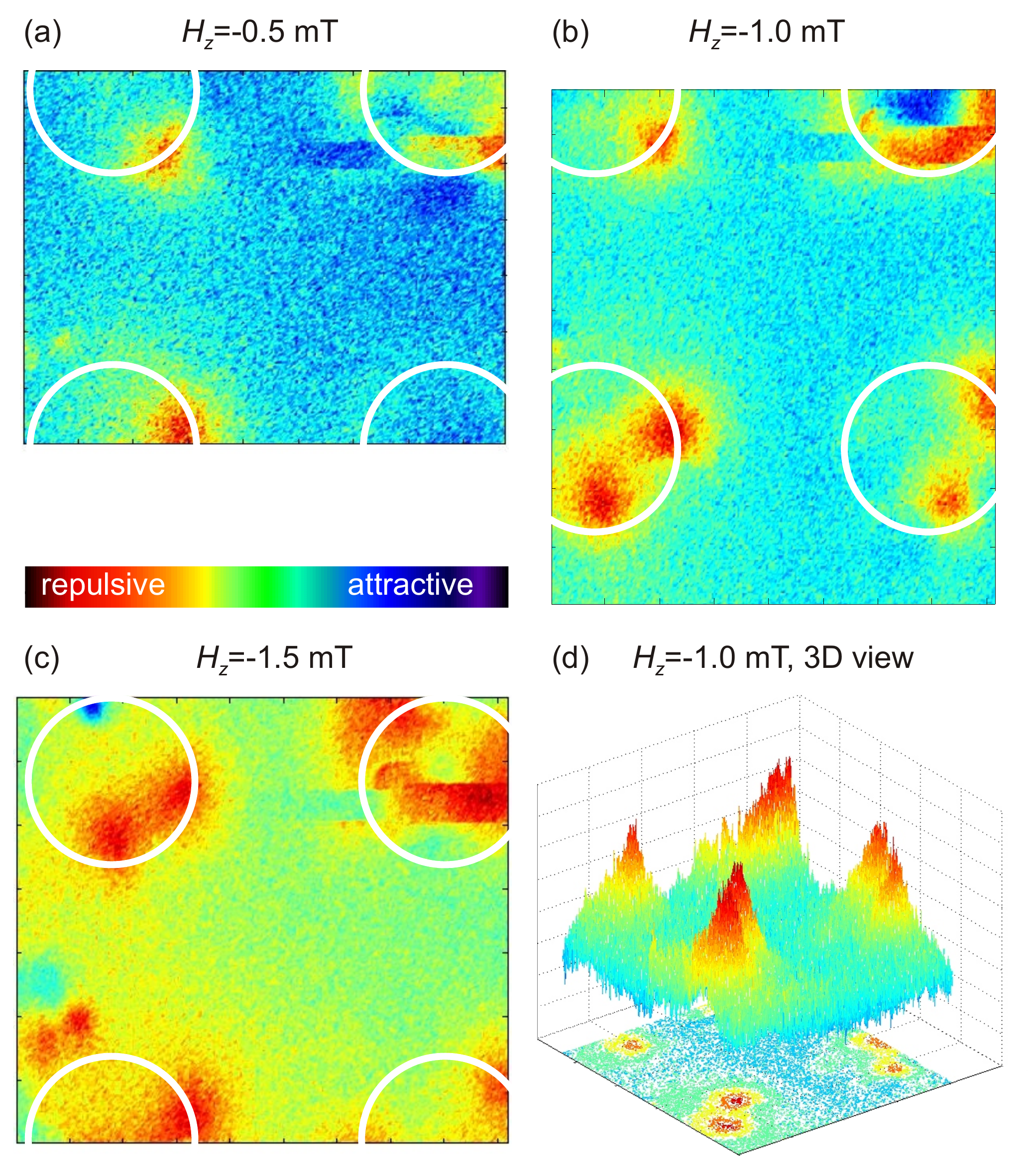}
		\caption{Visualization of superconducting vortices pinned by Py dots at 6.1\,K (72$\% T_c$). For a better visualization of the vortex positions the ``background'' image was subtracted. The frozen effective fields $H_z$ are: (a) $-0.5$\,mT, (b) $-1$\,mT and (c) $-1.5$\,mT. SC vortices are visualized as red spots. (d) 3D view of the image (b). According to the repulsive interaction between the tip and the SC vortex the vortices are imaged as hills.}
	\label{fig:MFM_field}
\end{figure}

An area where one dot is not fully switched to the magnetic-vortex state and has a residual in-plane component was chosen for LT-MFM imaging to ensure that the same dots are imaged at different fields and to correct a small thermal drift during experiments. 
It was established that the vertical coil of the microscope has a shift of zero point in the range of -0.5\,mT. This justifies to consider the $+0.5$\,mT image, where only magnetic contrast from the Py dots is observed, as a ``background'', and to subtract it from the other ones. The results are shown in Fig.\,\ref{fig:MFM_field} (a)--(c), respectively, and correspond to the effective fields $H_z = H_{\text{applied}} - 0.5$\,mT. The orientation of $H_z$ is negative, so that SC vortices and the MFM tip exhibit repulsive interaction and SC vortices show up as confined circular objects with positive frequency shift (red color). Hence, the SC vortices in Nb film have a polarity opposite to that of the magnetic-vortex core in Py dots. This means that the magnetostatic interaction between magnetic and SC vortices is repulsive. Such a configuration is selected to differentiate the magnetostatic pinning mechanism from the non-magnetic one. In the upper right dot, the residual in-plane component of the Py magnetization leads to a slightly disturbed difference image. 

Fig.\,\ref{fig:MFM_field}\,(a) corresponds to the first matching field $H_1$. Here about one SC vortex is visualized per unit area of the dot array, as expected. The SC vortices are located on top of the Py dots (white circles), showing that the dots work as preferable pinning centers. Nevertheless, they do not concentrate at the center of the dot, but occupy the edges of the dot. Furthermore, no SC vortices are visualized in the interstitial positions between Py dots. This effect becomes more pronounced when the second matching field $H_2$ has been applied during cooling [Fig.\,\ref{fig:MFM_field}\,(b)]. Also here, despite the long-range repulsive interaction between SC vortices, they are not distributed homogeneously, but are strongly pinned by the Py dots, so that two vortices are located on each dot. 
A further increase of the field to $H_3$ leads to an enhanced magnetic contrast on top of the Py dots, which corresponds to multiple flux quanta (vortex cluster) pinned by the dots [Fig.\,\ref{fig:MFM_field}\,(c)]. Here the expected three SC vortices could not be separately resolved due to overlapping of their magnetic stray fields at small vortex-vortex distances. The dark blue contrast that appears on one magnetic dot in this field (upper left dot) can be explained by the shift of the magnetic-vortex core by the stray field emanating from the observed vortex clusters. This dark blue contrast is stable and exists even at temperatures slightly above $T_c$ of the Nb film (image not shown here).

Based on the fact that two SC vortices are observed to be situated close to each other on top of the Py dot rather than being organized in a homogeneous Abrikosov lattice we can conclude that the pinning force at these artificial defects is higher than the repulsive force between vortices. In contrast to bulk superconductors where the vortex-vortex interaction force decays exponentially with the distance over the length of the penetration length $\lambda$, in thin films strong magnetostatic vortex-vortex interaction persist over long distance because the interaction occurs mainly outside the SC film without being screened by the SC~\cite{Bra09}. Consequently, the presence of a strong pinning potential is required to ensure the perturbation of the vortex lattice visualized here.

The maximal repulsive force between two SC vortices in thin films with a thickness below the penetration depth $\lambda$ can be approximated as
$F_\text{v-v} = \frac{\Phi_0^2}{\pi \mu_0 a^2}$,
where $a$ is the distance between the vortices~\cite{Pea64}. 
For the second matching field the distance between two vortices in the Nb film pinned by the same Py dot was measured to be about $a = 750$\,nm [Fig.\,\ref{fig:MFM_field}\,(b)].
Thus, the vortex-vortex repulsion force normalized by the Nb film thickness was estimated to be $F_\text{v-v} \approx 19\pm0.3$\,pN/$\mu$m. In reality, due to the finite size of the film, this force is slightly  reduced~\cite{Bra09}.
While scanning with the MFM tip, an additional force that acts on the SC vortices arises. This local tip-vortex interaction force can lead to a depinning of SC vortices and can be estimated from the MFM scans using the monopole-monopole model described in~\cite{Str08}. Using this model we have estimated the force that the MFM tip needs to exert on the supercondcuting vortex in order to depin it from its position. This force is found to be 1.5\,pN/$\mu$m for vortices in pure Nb film and 2.3\,pN/$\mu$m for the vortices located at the Py dots. Thus the pinning force of the vortices on the top of the Py dot, composed as a sum of the vortex-vortex repulsive force and tip-vortex dragging force, is estimated to be 21\,pN/$\mu$m, which is about 15 times stronger than the pinning force in the pure Nb film. The detailes of the pinning force analysis are described in Shapoval \textit{et al.}~\cite{Sha09}.  
 
On the one hand, our microscopic observations support the conclusion made from the magnetoresistance measurement that the Py dots in the magnetic-vortex state act as highly preferable pinning sites~\cite{Hof08}, on the other hand they show that a more detailed explanation of the pinning mechanism is essential for understanding the visualized arrangement of SC vortices in FM/SC hybrid structures. 
 
Two different mechanisms were proposed until now for the explanation of the enhanced pinning of SC vortices in the presence of a magnetic vortex. In the magnetostatic scenario, as it is described for example for an Al/Co hybrid structure~\cite{Vil08}, pinning occurs due to the magnetostatic interaction of the SC vortex with the magnetic vortex. In such a case, SC vortices with opposite polarity to the magnetic vortex are predicted to order themselves in interstitial positions of the dot array, whereas SC vortices with similar polarity should be located directly at the magnetic vortex core. Another mechanism is based on the local suppression of superconductivity due to the highly localized out-of-plane field produced by the magnetic-vortex core (core pinning)~\cite{Hof08}. Here, the magnetostatic interaction is negligible, and the SC vortices are located at the magnetic-vortex core independent of their polarity. As the observed SC vortices are localized at the edge of the Py dots, neither of these mechanisms can fully explain the situation reported in the present work.  

\begin{figure}[t]
	\centering
		\includegraphics[width=\columnwidth]{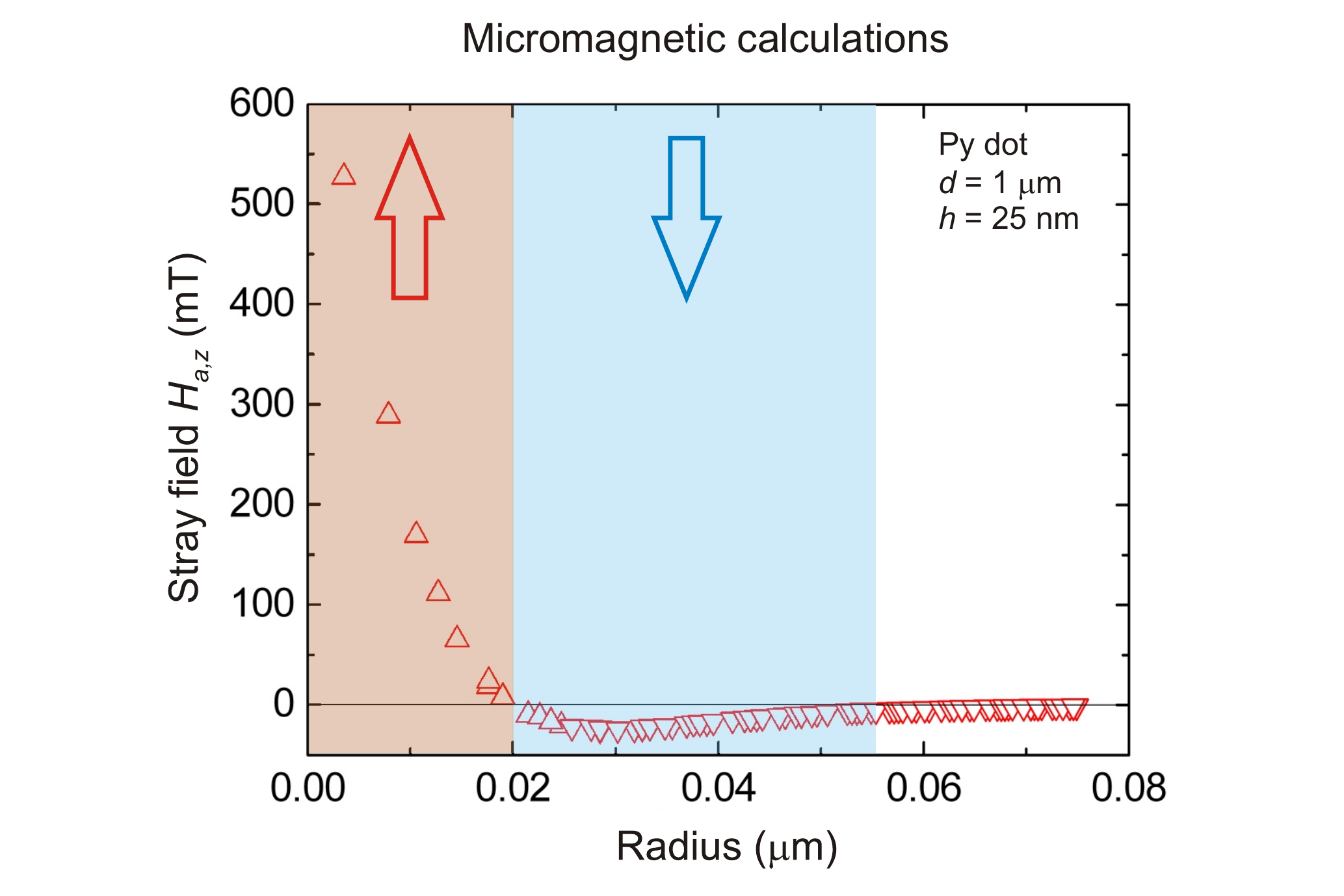}
		\caption{Stray field distribution just above the surface of the Py dot in the vortex state. The estimated values are based on micromagnetic calculations using the LLG-program~\cite{Wolf}. The magnetic-vortex core, where the strong out of plane component of the stray field comes out has a size of about 20\,nm. The main part of the dot has a zero out-of-plane component of the magnetization.}
	\label{fig:vortex_profile}
\end{figure}

To exclude the possibility that the SC vortices are magnetostatically attracted by the returning stray field lines of the magnetic vortex, the stray field distribution just above the surface of a Py dot in the magnetic-vortex state was estimated by micromagnetic calculations~\cite{Wolf}. The estimated stray field for the Py dot geometry used in the above-described experiments reveals that such negative field values are found already close to the distances of 25\,nm from the center of the dot (Fig.\,\ref{fig:vortex_profile}). Hence, the simple attractive magnetostatic interaction between the SC vortex and the returning stray field of the magnetic vortex could not provoke the visualized arrangement of the SC vortices being located almost at the edges of the Py dot [Fig.\,\ref{fig:MFM_field}\,(b)]. 

One should not forget that the magnetic stray field that comes out from the magnetic vortex decays with increased distance from the Py/Nb interface. Assuming that pinning is caused by the combination of the core pinning and magnetic repulsion bended SC vortices could occur in these hybride structures. The vortices pinned at the bottom of the Nb film by the magnetic vortex due to the core pinning mechanism can be bended and according to the repulsive interaction will shift to the edges of the dot. We can not fully exclude the possibility of this scenario, but due to the depth sensitivity of the MFM a bended SC vortex is expected to be imaged as a more elongated object.

A pinning mechanism which still depends on the presence of a highly permeable FM dots but is however independent of the polarity of the magnetic vortex, is the local polarization of the Py moments in the stray field of the SC vortex. This effect of the concentration of the magnetic flux in the presence of soft magnetic material may lead to the preferred positioning of SC vortices on top of the Py dots.   

Other important sources of the possible local enhancement of pinning that are independent of the magnetic properties of the Py dots should also be taken into account. In the present geometry of the hybrid structure the superconducting film is deposited on top of the Py dots. Thus, the existence of the 25\,nm thick Py dots underneath the Nb film leads to a surface modulation of the SC film. The AFM profile (image is not presented here) shows that the modulation $\approx 30$\,nm is on the scale of the Py dot thickness. This produces local stress in the SC film on the edges of the Py dots. These stressed areas can act as strong pinning sites and in such a way provoke the vortex distribution imaged in Fig.\,\ref{fig:MFM_field}. To check this fact nonmagnetic dots with the same geometry than the Py dots should be prepared. The behavior of the SC vortices in a Nb film that covers these non-magnetic dots should be compared with the presented results on FM/SC hybride structures. Moreover the intrinsic properties of the superconducting-normal metal interface could also lead to competitive pinning. 

In summary, we have demonstrated microscopically that the presence of magnetic vortices underneath the superconducting Nb film significantly influences the natural pinning landscape. The superconducting vortices are preferably located on top of the Py dots, experiencing a much stronger pinning force than SC vortices in the pure Nb film. This pinning force overcomes the repulsive interaction between the SC vortices, allowing SC vortex clusters be to formed and pinned by each dot. Our local magnetic force microscopic studies of the superconducting vortex distribution in the presence of an array of ferromagnetic dots with well controlled lateral dimensions are consistent with the global magnetoresistance measurements. Nevertheless, the reported local arrangement of the superconducting vortices could not be fully interpreted by the existing scenarios. Based on our experiments we propose two alternative explanations of enhanced pinning which do not depend on the polarity of the magnetic vortex: polarization of the soft magnetic dot and the stress induced pinning due to the surface corrugation. The main message from the local studies presented here is that conclusions about the pinning mechanism taken from global magnetization studies have to be judged very carefully and that further detailed local studies are required to understand the nature of pinning mechanisms in hybrids.  

\textit{Acknowledgements.} The work was partially funded by the European Community under the $6^{th}$ Framework Programme Contract Number 516858: HIPERCHEM and by the U.S.~NSF, grant ECCS-0823813. Image processing was done using the WSxM 4.0 software~\cite{Hor07}. The authors thank the ESF for financing a conference visit that resulted in a fruitful cooperation. The authors acknowledge discussions with E.~Reich and S.~Haindl, thank D.S.~Inosov and R.~Schäfer for critical reading of the manuscript and U.~Wolff for MFM tip preparation. 

\end{document}